\documentclass[journal]{IEEEtran}

\ifCLASSINFOpdf
\else
   \usepackage[dvips]{graphicx}
\fi
\usepackage{url}

\usepackage[pdftex]{graphicx}
\usepackage{blindtext}
\usepackage{xcolor}
\usepackage{dblfloatfix}
\usepackage{amsfonts}
\usepackage{amsmath}
\usepackage{mathrsfs,amsmath}  

\usepackage[inline]{enumitem}
\usepackage[pdftex]{graphicx}
\usepackage{amssymb}
\graphicspath{{../pdf/}{../jpeg/}}
\DeclareGraphicsExtensions{.pdf,.jpeg,.png}
\usepackage{subcaption}
\usepackage[utf8]{inputenc}
\usepackage[english]{babel}

\begin{document}

\title{Circular Convolution and Product Theorem for Affine Discrete Fractional Fourier Transform }

\author{Amir R. Nafchi, Eric Hamke \IEEEmembership{Member, IEEE}, Cristina Pereyra, and Ramiro Jordan, \IEEEmembership{Senior Member, IEEE}
\thanks{ A. R. Nafchi is with the University of New Mexico, Albuquerque, NM 87106 USA. (e-mail: raeisi@ieee.org).}
\thanks{E. E. Hamke, is with the University of New Mexico, Albuquerque, NM 87106 USA.(e-mail: ehamke@unm.edu)}
\thanks{C. Pereyra, is with the University of New Mexico, Albuquerque, NM 87106 USA.(e-mail: crisp@math.unm.edu)}
\thanks{R. Jordan, is with the University of New Mexico, Albuquerque, NM 87106 USA.(e-mail: rjordan@unm.edu)}}

\markboth{IEEE SIGNAL PROCESSING LETTERS, Vol. XX, No. XX,XXX XXX}
{Shell \MakeLowercase{\textit{et al.}}: Bare Demo of IEEEtran.cls for IEEE Journals}
\maketitle

\begin{abstract}
The Fractional Fourier Transform is an ubiquitous signal processing tool in basic and applied sciences. The Fractional Fourier Transform generalizes every property and application of the Fourier Transform.  Despite the practical importance of the discrete fractional Fourier transform, its applications in digital communications has been elusive. The convolution property of the discrete Fourier transform plays a vital role in designing multi-carrier modulation systems. Here we report a closed form affine discrete fractional Fourier transform and we show the circular convolution property for it. The proposed approach is versatile and generalizes the discrete Fourier transform and can find applications in Fourier based signal processing tools.
\end{abstract}


\IEEEpeerreviewmaketitle

\section{Introduction}
\IEEEPARstart{T}{he} discrete fractional Fourier transform (DFrFT) has been applied as an analysis tool in a variety of signal processing, radar and optics problems\cite{ncomms,wang2012sar,zhao2016parameter}. 
Although the continuous form of the transform, fractional Fourier transform (FRFT), has a closed form representation, yet there is no closed form representation of DFrFT that generalizes the discrete Fourier transform (DFT) and preserves its important properties.
\\Convolution and product property is an important feature of the DFT that enables applications in digital communications and design of multi-carrier modulation systems such as orthogonal frequency division multiplex (OFDM) \cite{zou1995cofdm}. 
An integral transform, $\mathcal{L}$, associates with the convolution operation, $\ast$, if it preserves the following property 
\begin{equation}  \label{eq_1_convolution}
\mathcal{L}(f \ast g) = \mathcal{L}(f) \mathcal{L}(g)
\end{equation} 
where $f$ and $g$ are $L^2$ functions.
\\ \indent Studying the convolution property for FrFT interested many researchers. Almeida derived the convolution and product properties for FrFT for the first time \cite{almeida}. 
Despite Almeida's efforts in deriving FrFT convolution and product property, this problem has been re-investigated by numerous researchers because of various reasons including complexity of methods or incompatibility with definition of aforementioned convolution operation \cite{wei2016,zayed2018two,zayed1998convolution,wei2019convolution,zhang2016new}.
\\ \indent Unlike DFT that inherits the convolution property from Fourier transform, DFrFT's convolution property is not easy to evaluate because of its kernel's quadratic phase components. Wei showed the convolution property for a uniformly sampled FrFT \cite{wei2016}. 
 The proposed convolution theorem considered infinite samples and do not hold for finite number of samples as we need for the definition of discrete circular convolution \cite{oppenheim}.
\\ \indent Discrete circular convolution property plays an important role in designing versatile multi-carrier modulation systems that can operate optimally under extreme channel conditions. The Orthogonal Frequency Division Multiplex (OFDM) is a famous example of such multi-carrier modulations that converts a frequency selective channel into a flat fading channel and owes this to circular convolution property of DFT \cite{stuber}. 
\\ Hence, discrete circular convolution property has a pivotal role in designing of multi-carrier  modulation systems and must be investigated for DFrFT so as to achieve a comparable system design.
\\ \indent In this paper we present an affine DFrFT and show convolution and product properties for it. 
\section{Affine Discrete Fractional Fourier Transform}
The affine DFrFT of a $L^2$  discrete signal $x[n]$ is defined as follows
\begin{equation}  \label{eq_2_DFrFT}
X_\alpha[k] = \mathscr{F}_\alpha\{{x}[n]\}[k]=\sum_{n=0}^{N-1}x[n]K_{\alpha}[k,n]
\end{equation} 
In (\ref{eq_2_DFrFT}), $\alpha$ is the rotation angle, in Radians, of transform  in time-frequency plane; $K_{\alpha}[k,n]$ is the kernel of the DFrFT given by
\begin{eqnarray}  \label{eq_3_kernel}
K_{\alpha}[k,n]= \begin{cases}
\kappa_\alpha e^{j2\pi\big (\frac{n^2+k^2}{2}{\rm cot}(\alpha)-\frac{nk}{N}\big )} \; \text{if $\alpha$ is not a multiple of $\pi$}\\
\delta(k - x) , \;\;\;\;\;\;\;\; \text{if $\alpha$ is a multiple of $2\pi$}
\\\delta(k + n) , \:\:\:\:\:\:\:\:\:\: \text{if $\alpha + \pi$ is a multiple of ${2}\pi$}
\end{cases}
\end{eqnarray} 
where $\kappa_\alpha$ in (\ref{eq_3_kernel}) is $\sqrt{\frac{1-j\,{\rm cot}(\alpha)}{N}}$, and $n,k\in \{0,...,N-1\}$.\\
The N-point DFrFT of signal $x[n]$ is expressed by
\begin{equation} \label{eq_4_mat}
 \mathbf{X}=W^{\alpha}\mathbf{x}  
\end{equation}
where $W^{\alpha}$ is the N-by-N square DFrFT matrix, and $\mathbf{x}$ is the $N \times 1$ vector representation of signal $x[n]$.

\begin{equation}
\label{DFrFTmat}
\resizebox{0.9\hsize}{!}{
$W^{\alpha}= 
\begin{pmatrix}
w_{0,0}^{\alpha} & w_{0,1}^{\alpha} & w_{0,2}^{\alpha}&w_{0,3}^{\alpha}&\cdots &w_{0,N-1}^{\alpha}\\ 
w_{1,0}^{\alpha} & w_{1,1}^{\alpha} & w_{1,2}^{\alpha}&w_{1,3}^{\alpha}&\cdots &w_{1,N-1}^{\alpha}\\ 
w_{2,0}^{\alpha} & w_{2,1}^{\alpha} & w_{2,2}^{\alpha}&w_{2,3}^{\alpha}&\cdots &w_{2,N-1}^{\alpha}\\ 
w_{3,0}^{\alpha} & w_{3,1}^{\alpha} & w_{3,2}^{\alpha}&w_{3,3}^{\alpha}&\cdots &w_{3,N-1}^{\alpha}\\ 
\vdots  & \vdots  & \vdots &\vdots &\ddots &\vdots  &\\ 
w_{N-1,0}^{\alpha} & w_{N-1,1}^{\alpha} & w_{N-1,2}^{\alpha}&w_{N-1,3}^{\alpha}&\cdots &w_{N-1,N-1}^{\alpha}\\
\end{pmatrix}$}
\end{equation}
\\where $w_{n,k}^{\alpha}$s are the elements of the matrix $W^{\alpha}$ at row $n$ and column $m$ and if $\alpha$ is not a multiple of $\pi$ is  
\begin{equation}
   w_{n,k}^{\alpha} = \kappa_\alpha e^{j2\pi\big (\frac{n^2+k^2}{2}{\rm cot}(\alpha)-\frac{nk}{N}\big )}
\end{equation}
For $\alpha= \frac{\pi}{2}$ it becomes the DFT matrix. Unlike DFT the DFrFT matrix, $W^{\alpha}$, is not circulant. Circular convolution property of DFT stems from circulant property.

\section{Circular Convolution for Discrete Fractional Fourier Transform}
This section examines circular convolution property for the suggested affine DFrFT.\\
\textbf{\textit{Definition 1:}} For finite sequences $x[n]$ and $h[n]$ of length $N$ the circular convolution is
\begin{equation}  \label{eq_circularConv}
y[n]=(h\circledast_N x)[n]=\sum_{m=0}^{N-1}h[m]x[(n-m)_N]
\end{equation} 
where $n$ belongs to $[0,..,N-1]$, and $(n-m)_N$ is the modulo operation. 
\\ \textbf{\textit{Definition 2:}} We define $\tilde{y}$, $\tilde{x}$ as:
\begin{align}   \label{eq_tilde}
\tilde{x}[n]&=x[n]e^{An^2}\\
\tilde{h}[n]&=h[n]e^{An^2}  
\end{align}

{In \eqref{eq_tilde}, we define $A$ as follows
\begin{equation} \label{A_coeff}
A=j\pi\,{\rm cot}(\alpha ).
\end{equation}}
Now we define chirp-circular convolution 
\begin{equation}  \label{eq_ChirpCircConv}
\tilde{y}[n]=\kappa_\alpha \rm (\tilde{h}\circledast_N \tilde{x})[n]\,e^{-An^2}
\end{equation}\\
We will prove that the affine transform of $\tilde{y}$ is:

\begin{eqnarray} \label{eq_DFrFT_circ}
Y[k]&=&\mathscr{F}_\alpha\{\tilde{y}[n]\}[k]\\ \nonumber&=&{\kappa_\alpha}^2 \rm \sum_{n=0}^{N-1}(\tilde{h}\circledast_N \tilde{x})[n]\,e^{-An^2}e^{A(n^2+k^2)+Bnk}\\ \nonumber&=&H[k]X[k]e^{-Ak^2}
\end{eqnarray}
where $B=-j\frac{2\pi}{N} $.

In above equation $H[k]$,$X[k]$ are the fractional transforms of $h[n]$ and $x[n]$ respectively.
\\ \textbf{\textit{Proof:}} 
\begin{eqnarray}  \label{eq_proof1}
Y[k]&=&{\kappa_\alpha}^2 \rm\sum_{n=0}^{N-1}(\tilde{h}\circledast_N \tilde{x})[n]\,e^{-An^2}e^{A(n^2+k^2)+Bnk}\\\nonumber
&=&{\kappa_\alpha}^2 \rm \sum_{n=0}^{N-1}\Bigg( \sum_{m=0}^{N-1}h[m]e^{Am^2}x[(n-m)_N]e^{A((n-m)_N)^2}  \Bigg)\\\nonumber & \times &e^{-An^2}e^{A(n^2+k^2)+Bnk}
\end{eqnarray} 
From the definition of modulo arithmetic we can write 
\begin{eqnarray}  \label{eq_mod}
{(n-m)}_N = \begin{cases}
n-m , &\text{if $n-m\geq 0$}\\
n-m+N ,  &\text{if $n-m < 0$}
\end{cases}
\end{eqnarray}
By replacing equation \eqref{eq_mod} in  \eqref{eq_proof1} we will have: 
\begin{eqnarray} \nonumber  \label{eq_proof2} 
Y[k]&=&{\kappa_\alpha}^2 \rm\sum_{m=0}^{N-1}h[m]e^{Am^2}\Bigg( \sum_{n=m}^{N-1}x[(n-m)]e^{A(n-m)^2}e^{Bnk} \\&+& \sum_{n=0}^{m-1}x[(n-m+N)]e^{A(n-m+N)^2}e^{Bnk} \Bigg)\\\nonumber & \times &e^{Ak^2}
\end{eqnarray}
By taking $n-m=p$, and $n-m+N=p$, equation \ref{eq_proof2} becomes: 
\begin{eqnarray} \nonumber  \label{eq_proof3} 
Y[k]&=&{\kappa_\alpha}^2 \rm \sum_{m=0}^{N-1}h[m]e^{Am^2}\Bigg( \sum_{p=0}^{N-m-1}x[(p)]e^{A(p^2+k^2)+Bk(p+m)} \\&+& \sum_{p=N-m}^{N-1}x[(p)]e^{A(p^2+k^2)+Bk(p+m-N)} \Bigg)\\\nonumber
\end{eqnarray}
Then (\ref{eq_proof3}) becomes
\begin{eqnarray} \nonumber  \label{eq_proof3} 
Y[k]&=&{\kappa_\alpha} \rm \sum_{m=0}^{N-1}h[m]e^{A(m^2)+Bkm} \\ \nonumber &\times& \kappa_\alpha \sum_{p=0}^{N-1}x[p]e^{A(p^2+k^2)+Bkp}\\\nonumber
&=&\Bigg( \kappa_\alpha \sum_{m=0}^{N-1}h[m]e^{Am^2+Bmk}\Bigg)X[k]\\ \nonumber
&=&H[k]X[k]e^{-Ak^2}\\
\end{eqnarray}
This completes the proof of (\ref{eq_DFrFT_circ}). The result of the transform in fractional domain contains phase shift term $e^{-Ak^2}$. For filtering purposes, given knowledge of channel the $X[k]$ can be recovered with a single tap equalizer in fractional domain.

\section{Concluding Remarks}
This letter presented a novel affine DFrFT that preserves circular convolution property. The proposed transform is versatile and is applicable in various areas that DFT is an indispensable tool. Future work will include designing communications application based on the proposed affine DFrFT transform. It can find applications in designing multi-carrier modulation systems such as orthogonal frequency division multiplex.


\bibliographystyle{unsrt}
\bibliography{bare_jrnl}
\end{document}